\documentclass[aps,pra, onecolumn,14 pts,a4paper,
 ]{revtex4}

\usepackage{amssymb,amsmath,amsfonts,graphicx}
\usepackage{bm}
\usepackage{times}
\usepackage{graphicx}
\usepackage{bm}
\usepackage{color}

\definecolor{pink}{rgb}{1,1,0} 
\definecolor{red}{rgb}{1,0,0}
\definecolor{yellow}{rgb}{1,1,0}
\definecolor{orange}{rgb}{1,0.5,0} 
\definecolor{blue}{rgb}{0,0,1}
\definecolor{white}{rgb}{1,1,1}

\begin{document}

\title{Is vacuum dispersive?}

\author{Yves Pomeau}
\affiliation{ Department of Mathematics, University of Arizona, Tucson, USA}

\date{\today }
\begin{abstract}
\textbf{Abstract}
The question we ask is: does the speed of light  {\it{in vacuo}} depend on its frequency? While the answer is NO in the frame of classical physics, we point out that the opposite could be  true if one take into account the polarization of Dirac sea. We  estimate the dependence of the index of refraction of vacuum + Dirac sea versus the wavelength of an incoming beam, and suggest a way to test this effect.

\end{abstract}

\maketitle

Is vacuum dispersive? In other terms, does the speed of light  {\it{in vacuo}} depend on its frequency? The answer is obviously NO if one considers a classical (= non quantum, namely with many photons in the same state) electromagnetic wave (EM wave later - this "EM" will be attached either to the incoming wave or the zero point fluctuations, hopefully no confusion will arise) described mathematically by Maxwell equations \cite{maxwell} {\it{in vacuo}}. This assumes a "geometric" vacuum  supporting perturbations propagating 
at the speed of light and symmetric under the Poincar\'e group. Even by making abstraction of general relativity, developments in physics since Maxwell have shown that what we call vacuum is endowed with other properties than what Maxwell assumed. Among them is the fact that this so-called vacuum is filled with  quantum fluctuations. As discovered by Planck \cite{planck} the ground state of the EM field is subject to zero point quantum fluctuations, that have been shown since to have measurable consequences. The Casimir pressure \cite{casimir} between two parallel conducting plates as well as other related effects are observable consequences of zero point fluctuations of the EM field {\it{in vacuo}}. But those zero point fluctuations of the EM field have no effect on the dispersion of a propagating  EM wave, because they do not interact at all with this wave,  due to the linearity of the wave equations.  

There is another source of quantum fluctuations {\it{in vacuo}}, the Dirac sea. In Dirac theory \cite{dirac} of the quantum relativistic electrons, the ground state is such that each state of negative energy is occupied by one electron, all of them making the Dirac sea. Those electrons should interact with a propagating EM wave just because of the structure of Maxwell and Dirac equations. In Dirac equation for an electron in a given EM field this interaction depends on three dimensionless numbers. One is the ratio  
 $$r(\omega) =  \frac{\hbar \omega}{(m_e c^2)} \mathrm{,}$$ $\omega$ pulsation of the propagating EM wave, $\hbar = 2\pi$ times Planck's constant, $m_e$ electron mass and $c$ speed of light. For an EM wave in the visible range $\frac{\hbar \omega}{(m_e c^2)}$ is a small but not very small number (related in a simple way to the fine structure constant if $\omega$ is the pulsation of the light emitted by a transition of a Hydrogen atom), so that this interaction depends on $\omega$ and of the way (and the exponent) this dimensionless combination appears in the formula for dispersion for $\omega$ small.  
 
 The other dimensionless number describing the interaction between EM waves and electrons and therefore with the Dirac sea is the fine structure constant $\frac{e^2}{\hbar c}$, an universal and small constant ($e$ charge of the electron). There should be a third constant representing the amplitude of the EM wave, compared to fundamental quantities like the rest energy of an electron. However, because we have to deal with the linear approximation only when considering the index of refraction, this amplitude-dependent constant disappears in the formula for this index of refraction.

 Consider how the refractive index $n$ of vacuum, a dimensionless quantity, could  depend on the  frequency of the EM wave. It should be a numerical function of $r(\omega)$ and of the fine structure constant. It cannot depend on the sign of $\omega$, itself a function of the time direction, irrelevant for the real part of the refractive index we consider. Therefore $n$ should be an even function of $r(\omega)$. In the limit where this quantity is small it is natural to approximate  $n(.)$ by the first two terms of its Taylor expansion, with a term quadratic with respect to $r(\omega)$ after the first one:   
 $$ n \approx 1 + k \left(\frac{\hbar \omega}{m_e c^2}\right)^2 \mathrm{,}$$ where $k$ is a function of the fine structure constant. The simplest dependence for $k$ is that it is proportional to the fine structure constant. Because this polarization term comes from the  interaction between charged electrons in the Dirac sea and a propagating EM wave, one finds a first power of $e$, as it enters in the interaction between the EM field and an electron as described by the Dirac equation. There is another factor of $e$ because once one has derived the perturbation brought to the quantum states in the Dirac sea, one has to derive from it a dipolar moment which yields another factor $e$, whence a factor $e^2$ in the contribution to the correction to the index of refraction and lastly a contribution proportional to the fine structure constant which is quadratic in $e^2$.  We outline below the principles of a calculation of this effect, but for the moment it is enough to give an expected order of magnitude of the dispersion for waves in the visible range. The order of magnitude of the dispersive part of the index of refraction is $$k' \frac{e^2}{\hbar c} \left(\frac{\hbar \omega}{m_e c^2}\right)^2 \mathrm{,}$$
 where $k'$ is a pure number expected to be of order one. For frequencies in the visible range $r(\omega) =  \frac{\hbar \omega}{(m_e c^2)}$ is a fraction of the fine structure constant. Therefore a very rough estimate for the value of the correction to the index of refraction due to the polarization of the Dirac sea is a small fraction of the fine structure constant to the cube, something in the range $10^{-8}$ to $10^{-9}$, far too small for having been seen in measurements of the speed of light on Earth, having a relative accuracy of order $10^{-6}$. However, as explained below, we believe that this dispersion effect could be measured or at least looked at in astrophysical conditions.  
 
  A way to measure a possible frequency dependence of the speed of propagation of EM waves {\it{in vacuo}} would be to adapt a presently working device. This device measures the Earth-Moon distance by shooting a powerful laser pulse toward a mirror sitting on the Moon, detecting return photons and measuring the back and forth travel time \cite{fourbinicois}. The idea would be to shoot {\it{two}} beams (at least) at exactly the same time (or with a well controlled small time difference) in the same direction but with two different frequencies of light and to measure a possible difference of back and forth travel time between the two beams, something that has not been done so far to the best of our knowledge. The two beams could have the same source, with one beam at the double frequency of the other after traversing a frequency doubling cell. In the device described in reference  \cite{fourbinicois}, such frequency doubling is done in the laser cavity but it is not said if the photons detected in the return beam have the undoubled or doubled frequency. 
  The accuracy of this device is amazing: the Earth-Moon distance is said to be measured within a few millimeters (this is the distance between the laser on Earth and mirrors left on the Moon by Space missions). The travel time is about 600 seconds and is known with an accuracy of $60$ picoseconds. This yields a relative accuracy of $10^{-13}$, largely sufficient to detect a variation of the index of refraction of the space between Earth and Moon of order $10^{-9}$, the order of magnitude given above.  Let us also notice that this neglects other possible sources of dispersion, like the existence of a plasma (in the usual meaning) due to the solar wind or the perturbation to the Dirac sea due to magnetic fields from the Earth and the Sun or the difference of travel time across the atmosphere between the two colors of light. 
  
 Let us sketch a theoretical approach of the possible dispersion of EM waves by interaction with the Dirac sea. This could be done as follows. Each (negative) energy state of the Dirac sea is described by a plane wave solution of Dirac equation: The perturbation brought by the EM wave to Dirac equation is represented by the addition of $(-eA)$ to the four component momentum operator, $A$ being the potential vector of the EM wave. Another correction to the wave function could be needed to ensure that the perturbed wavefunction of the full Dirac sea remains anti-symmetric. Once the perturbation to the plane wave solution of Dirac equation is known, one should derive from it the polarization of this perturbed state, the polarizability of Dirac sea and so the change in the index of refraction due to it. There are (at least) two possible causes of failure of this program. First the final result for the polarizability may diverge because either the low or large energy electrons, or both. Another possible cause of failure is not related directly to a computational problem, but to physics: the polarization of Dirac sea is constrained by what we know of the physical world: this polarization cannot diverge at zero frequency, the vacuum plus the Dirac sea making obviously an insulator in our world. Therefore the polarizability of Dirac sea at vanishing frequencies has to be at least finite or zero (which is better). This may be very constraining because if one views Dirac sea as a kind of dense plasma, the polarizability of such a plasma diverges at zero frequency (a way of telling that the plasma conduct electricity at zero frequency). A coming publication will be devoted to those questions. 
 
 As a last remark, a dispersive vacuum means, among other things, that this vacuum defines an ether, namely that not all Lorentz frames are equivalent. Changing the frequency of an EM wave can be done by a Lorentzian boost so that, if vacuum was dispersive, the speed of propagation of EM waves depends on the frame of reference. 

      {}

      \ifx\mainismaster\UnDef%
      \end{document}
      \fi